\documentclass[12pt,preprint]{aastex}

\begin{document}

\title{Line and Mean Opacities for Ultracool Dwarfs and Extrasolar
Planets}

\author{Richard S. Freedman\altaffilmark{1} \& Mark S. Marley}

\affil{NASA Ames Research Center}

\affil{Mail Stop 245-3}

\affil{Moffett Field, CA 94035-1000}

\email{freedman@darkstar.arc.nasa.gov}

\author{Katharina Lodders}

\affil{Planetary Chemistry Laboratory, Department of Earth and Planetary
Sciences, Washington University, Campus Box 1169, Saint Louis, MO
63130-4899}

\email{lodders@wustl.edu}

\altaffiltext{1}{SETI Institute, Mountain View, CA}

\begin{abstract}
Opacities and chemical abundance data are crucial ingredients of
ultracool dwarf and extrasolar giant planet atmosphere models.  We
report here on the detailed sources of molecular opacity data employed by our
group for this application.  We also present tables
of Rosseland and Planck mean opacities which are of use in some studies
of the atmospheres, interiors, and evolution of
planets and brown dwarfs.
For the tables presented here we have
included the opacities of important atomic and molecular species, including the alkali elements, 
pressure induced absorption by hydrogen,  and other significant opacity
sources  but neglect opacity from condensates.  We report for each species how we have assembled
molecular line data from a combination of public databases, laboratory data
that is not yet in the public databases, and our own numerical calculations.
We combine these opacities with abundances
computed from a chemical equilibrium model using recently revised solar abundances
to compute mean opacities.  The chemical equilibrium calculation
accounts for the settling of condensates in a gravitational field, and
is applicable to ultracool dwarf and extrasolar planetary atmospheres, but not
circumstellar disks.  We find that the inclusion of alkali atomic opacity
substantially increases the mean opacities over those currently in the literature
at densities relevant to the atmospheres
and interiors of giant planets and brown dwarfs.   We provide our opacity
tables for public use and discuss their limitations.

\end{abstract}


\section{Introduction}

The thermal structure and radiative transfer within the atmospheres of
ultracool dwarfs and extrasolar planets ultimately depends upon an
entire suite of molecular and atomic opacities relevant to the the
temperatures and pressures found in these objects.  Our group utilizes
these opacities in models of ultracool dwarfs and extrasolar planets
\cite[e.g.,][]{Marley et al. 2002, Fortney et al. 2006, Saumon et al. 2007} and provides them to other members
of the community.   Here we briefly summarize our current suite of
molecular opacities that we depend upon in our modeling and consider
some relevant issues in their construction. We also report on the
chemical equilibrium calculation used to compute molecular abundances.

Although less commonly used than in the past, some calculations,
particularly of planetary evolution, utilize Rosseland ($\kappa_{\rm R}$) or Planck ($\kappa_{\rm P}$) mean
opacities \citep[e.g.,][]{Hubickyj et al. 2005}.   Canonically, Rosseland
mean opacities are the appropriate choice for radiative transfer models
of optically thick atmospheric regions where radiation propagates by
diffusion, whereas Planck mean opacities are the appropriate choice in
optically thin regions \cite{Mihalas 1978}.  
By their nature, mean opacities are sensitive to the sum of the relevant opacity sources involved in their computation.  
Changes both in molecular and atomic abundances and in the individual opacity sources themselves can impact the final product.  Here
we present improved calculations for both the abundance of atoms used in our equation of state and new calculations
for the relevant opacity sources.

The best available data defining the `solar abundance' of the atoms has evolved
over time as understanding of the sun's atmospheric composition has improved.  Here we employ
solar and scaled solar metallicities that use the recently revised
abundances presented in \cite{Lodders 2003}. The chemical equilibrium gas
compositions computed as a function of temperature and total pressure
take condensate formation and condensate settling into clouds into
account in the calculation of the chemical equilibrium as is appropriate for the conditions in brown dwarfs, 
extrasolar giant planets and cool stars.  However we purposefully neglect grain opacity in the calculation
of the Rosseland and Planck means.  We emphasize that the results here
do not apply to lower-gravity environments such as proto-planetary
disks because there the condensation chemistry, and therefore the gas
chemistry, is significantly different than that in gravitationally
bound atmospheres of substellar objects \cite{Lodders 1999a,  Lodders and Fegley 2002, Lodders and Fegley 2006}.  

To compute mean opacities, we calculated a new set of molecular and atomic
opacities on a grid of 324 pressure and temperature points
ranging from 75 to 4000K and $3\times 10^{2}$ to $3\times 10^{8}\,\rm dyne\,cm^{-2}$
($3\times 10^{-4}$ to 300 bar).   Every atomic and molecular line opacity in our database 
of 10 different gaseous species was included, and not simply a sampling.  These opacities
were computed on a fixed wavenumber grid that completely resolved individual line profiles.
These opacities were then combined with
collision induced absorption due to interactions of hydrogen and helium
as well as several other opacity sources. This new set of mean
opacities uses the latest physical information including new terms due
to alkali atoms that were not included in previous investigations. Condensate opacity is not
included in the mean opacities presented here because condensate opacities
depend on the particulars of cloud models chosen to model substellar
atmospheres \citep[e.g.,][]{Ackerman and Marley 2001} and may very
greatly, depending upon the actual thermal structure in a given
situation.  Because we neglect grain opacity, the opacities presented here should
be regarded as a lower limit to the true opacity, which may be several orders of magnitude higher.  
The grain free results are of 
interest as they highlight the important role of the alkali elements and isolate
assumptions regarding the grain opacity.

Recent tabulations of mean opacities include the work by Lenzuni et al. (1991)
for a zero metallicity gas and Ferguson et al. (2005).  Tabulations from the latter work
generally  include opacity of solid condensates but a few cases without grains
are also presented.  We compare our results to the latter work
over the relatively limited region of overlap.  No other pure gaseous opacity tables
are readily available for solar metallicity over our temperature and pressure regime.

We discuss the opacity sources and methods for treating line broadening
in Section 2.  In Section 3 we
discuss our chemical equilibrium calculation.  Mean opacities are
presented and discussed in Section 4.

\section{Molecular Opacity Sources and Data}

We maintain a large and constantly updated database of molecular and
atomic opacities.  Some of the opacities
are from standard sources, such as the HITRAN database, while others
are a mixture of standard and other sources.  In this section we discuss
the opacity sources we employ for each molecule of interest.  Sharp and Burrows (2007)
recently published a thorough discussion of molecular opacities for
ultracool dwarfs and discuss which species are of greatest interest for modeling
these objects.  We refer the reader to that work for more extensive background discussions than
are included here.  Furthermore,  several of the molecular and atomic opacity
tabulations that they review are also employed by our group.  Thus for a number
of opacity sources we simply defer to their discussion.  In other
cases, particularly for $\rm CH_4$, our opacity line list is unique and
we discuss it in some detail. 

Given a list of atomic transitions, it is further necessary to
compute a line shape for each atomic or molecular line.  Thus in
Section 2.2 we discuss our choices for molecular line shapes and pressure
broadening.

\subsection{Molecular opacities}

In this section we discuss the molecules for which we compute opacities
(although many more species are included in the
chemical equilibrium calculation).  The selection of molecules is
dictated by the chemistry in solar-composition brown dwarf and 
extrasolar giant planet atmospheres, and, to some extent, the availability
of line data.  Our list includes the most important opacity sources at
the temperatures appropriate to our calculations as validated by
observations of cool stars, brown dwarfs, and giant planets. We do not include certain
sources that affect only the far UV portion of the spectra as our high
temperature cutoff is currently 4000 K.

Most spectroscopic databases are built upon measurements taken at or
near room temperature, and theoretical calculations supply the missing
transitions that can become important at high temperatures. If only
room temperature databases are used, the transitions from highly
excited energy levels (usually referred to as hot bands in the
literature) would be missing and the true opacity at elevated
temperatures would be substantially underestimated. Whenever possible,
we use expanded databases here.  

Note that all references to HITRAN are to the HITRAN database. The HITRAN
website\footnote{ \url{http://cfa-www.harvard.edu/hitran}} provides the
latest, updated data and copies of all the papers
that give details on each molecule included.

\begin{itemize}

\item $\rm H_{2}O$:  The two most extensive line lists for water are
those of Partridge and Schwenke (1997) and Barber et al. (2006).  Both
are computational lists with hundreds of millions of lines.  We have
made extensive comparisons between the two lists and find that over the
temperature range relevant to brown dwarf and extrasolar planet
atmospheres, the differences are slight.  We utilize the former list
here, supplemented with lines for minor isotopes ($\rm HD^{18}O$ and
$\rm HD^{17}O$) from HITRAN\footnote{In calculations specifically for
brown dwarfs we set the deuterium abundance equal to zero, see \S 3.1.}. The
entire database includes about $2.9\times10^8$ lines.  Line widths are
computed using $\rm H_2$ broadening data from Gamache \citep{Gamache et
al. 1998, Gamache 2001};  a recent paper 
\cite{Ma et al. 2007} discusses possible problems with the current
implementation of this theory for some other molecules. However, the few examples shown in the Ma et al. paper only cover HF
broadened by HF or $\rm N_2$.  The data for $\rm N_2 - HF$ broadening showed little differences
from the earlier theory, leading us to have some confidence in our approach until data is available
for $\rm H_2$ broadening of $\rm H_2O$.  

We note that Allard et al. \cite{Allard et al. 2000} considered the completeness
of the Schwenke H$_{\rm{2}}$O (and TiO) databases at high temperature
and
concluded that the water database still lacked transitions from high
vibrational energy levels needed for calculating models of M stars and proposed
that a then preliminary (and never, to the best of our knowledge, publicly released) database by the Tennyson group was superior for these purposes.  
However, as discussed above the latest release of the
H$_{\rm{2}}$O database from the Tennyson
group \cite{Barber et al. 2006} no longer shows a significant difference with the earlier Schwenke
result at these temperatures.

\item $\rm CH_4$:  The laboratory analysis of the methane spectrum is
incomplete and is unlikely to be completed from laboratory measurements alone.  The
difficulties arise from the high
degree of symmetry of the molecule, which causes a great number of the bands
to overlap, and the fact that the average line separation is generally less than the
doppler width at room temperature.

The HITRAN methane database is not a true high temperature
database because many bands originating from higher vibrational levels
are not included.  Instead our line lists for
$\rm ^{12}CH_{4}$ and $\rm ^{13}CH_4$  were generated using the latest Spherical Top Data System (STDS)
software \cite{Wenger and Champion 1998} from the group at the University of
Bourgogne\footnote{\url{http://icb.u-bourgogne.fr/OMR/SMA/SHTDS/STDS.html}} which allows
us to calculate methane spectra to much higher total angular momentum
value, $J$, than tabulated in HITRAN.   For the calculations reported
here we computed to a maximum rotational level of
$J_{\rm max}=60$, which likely covers
lines of importance in cool objects. Homeier et al.  \cite{Homeier et al. 2003} also used
the STDS software, but only calculated up to $J_{\rm max}=40$.
For our calculation  one band of $\rm ^{13}CH_4$ that is missing in the
Bourgogne list was added from HITRAN and $\rm CH_3D$ was included from
HITRAN. Extensive line width data for all bands is not available
so we separated the available information by
reference to the vibrational symmetry of the transitions and applied it
to all bands \citep{Brown 1996}. There
are about $2\times10^8$ total $\rm CH_4$ lines in the database. Figure 1
compares our
computed hot methane opacity at 1000 K and $10^6\,\rm dyne\,cm^{-2}$ to an opacity computed solely from HITRAN data,
both without any $\rm CH_3D$.
There is generally good agreement near band centers and the computational database clearly provides
important additions in regions where the HITRAN data are lacking. 

Both our database generated from the STDS software and HITRAN
do not extend to wavenumbers much higher than about 6400 $\rm cm^{-1}$ (although
many weak lines due to the $\rm 2\nu_{3}$ band extend up to 10000 $\rm cm^{-1}$).
We thus supplement the numerical line database with a continuum opacity derived from
laboratory data by Strong \citep{Strong et al. 1993} even though the experimental results are not available at
elevated temperatures. This situation is unlikely to improve until new
theoretical predictions are available for $\rm CH_4$. The currently available
laboratory data for bands above 6400 $\rm cm^{-1}$ remained unanalyzed, in part because
of lack of knowledge of the detailed energy levels of the transitions.

Recently Leggett et al. (2007) compared our model spectra produced with the methane line list described above
to the  spectra of late type T dwarfs.  They found that the models still provide a poor match to observed spectra near $1.67\,\rm \mu m$, where the models show too much flux.
This is just the region where the calculated hot methane line list effectively
ends except for the few very high $J$ lines that extend to shorter wavelengths.  We note that
the CIA opacity of $\rm H_2-H_2$ (see below) is predicted to vary rapidly with wavelength (by about a factor of 2)
in this region. Thus, the inability to match the observations in this region could
be ascribed either to the lack of good methane data, problems in the CIA simulations,
or some combination of the two factors. It is also conceivable, but unlikely, that some other opacity source could also
be missing from the models in this region.  Further improvements to the  methane opacity
data will eventually resolve this problem.

\item $\rm NH_{3}$: For ammonia we rely on the line list from HITRAN 
supplemented with additional measurements, not yet fully analyzed, made at room temperature. 
 These additional lines are in the
6600 - 7000 $\rm cm^{-1}$ region and are from the $\nu_3 + 2\nu_4$,
$\nu_1 + \nu_3$, and the $2\nu_3$ bands \citep{McBride and Nicholls 1972, Brown 2000}.
Only line strengths at $296\,\rm K$  and estimated values
for the lower energy level are available for these bands.
The final line list has about
34,000 lines. Line widths are computed as arising 90\% from collisions
with $\rm H_2$ and 10\% with He (from work by Nemchikov as reported in Brown (2000)).

In their comparison of our spectral models for late T dwarfs to data, Leggett et al. (2007) also considered exploratory
models that employed laboratory measurements of ammonia opacity by Irwin et al. (1999) over the spectral range
of 0.91 to $1.9\,\rm\mu m$.  They found that even when atmospheric depletion of ammonia by
non-equilibrium chemistry was accounted for, the predicted near-infrared ammonia features were not seen
in the spectra of two T8 dwarfs.  We conclude that the Irwin et al. data overestimate the ammonia opacity and
we do not employ that dataset here.

\item CO: As with methane, for carbon monoxide we favor a high
temperature line list over that available from HITRAN. We utilize the
list from Goorvitch \cite{Goorvitch 1994}  which includes bands that originate
from highly excited energy levels.  We supplement this list with data
on $\Delta V=4$ transitions from Tipping (1993) and somewhat
fragmentary information on $\rm H_2$ and He line widths from the
literature (Bulanin et al. 1984, Le Moal and Severin 1986, Mannucci
1991).  Minor isotopes missing from the Goorvitch list were added from HITRAN.

\item $\rm  H_2S$: Data for the main isotope of $\rm H_{2}\,^{32}S$ are
from a calculated list by Richard
Wattson \cite{Wattson 1996} plus minor isotopes from HITRAN. $\rm H_{2}$
broadening was included from data in the literature \cite{Kissel et
al. 2002}. There are about 188,000
lines in the list. This list is not a true high temperature list but
does include
many weak lines below the intensity cutoff of the current HITRAN line
list.
The wavenumber coverage is thus much greater than HITRAN, extending to
19500 $\rm cm^{-1}$.

\item $\rm PH_{3}$:  For this molecule we use the latest HITRAN list,
including new bands and broadening information
from Linda Brown \cite{Brown 2000}. The list includes about 20,000
lines.

\item TiO: We include five isotopes  in our TiO tabulation from
David Schwenke \citep{Schwenke 1998} with modifications to the
strengths of the $\delta$ and $\phi$ bands
based on a comparison \cite{Allard et al. 2000} of models with
stellar spectra.
The list of about $1.7 \times 10^8$ lines includes transitions from
higher energy levels. There is no data on broadening from $\rm H_2$,
instead we compare data from other species such as $\rm H_2S$ to try to
set some reasonable limits on the broadening as discussed in \S 2.2.
The TiO molecular opacities should be reasonably complete
for all temperatures  considered except perhaps for the very highest
values considered in these tables \citep{Allard et al.
2000}

As reported in Sharp and Burrows (2007), one of us (RF) discovered an error in line
strengths in the program to convert the predictions of Schwenke
\cite{Schwenke 1998} from atomic units to HITRAN units,
with the strengths being off by a multiplicative factor of $2J^{\prime
\prime} +1$. This error is corrected here.

\item {VO:} For this molecule we rely on the line list, consisting of
about 3.1 million lines, from Plez (1998) which is briefly reviewed in
Sharp and Burrows (2007).  No information on line broadening is
available, so it is treated as was $\rm TiO$. 

\item {FeH:} The line list for FeH is known to be incomplete in the near infrared \cite{Cushing
et al. 2005}.  We rely on lists tabulated by
Dulick et al. (2003) and discussed in more detail by Sharp and Burrows (2007).  Only the most abundant $\rm ^{56}Fe$  isotope is
used.  No width data are available so it was estimated by using data for similar molecules from the literature.  

\item  {CrH:} For this molecule, primarily of interest in M and L
dwarfs, we rely on a list from Burrows et al. (2002) which is further reviewed by \cite{Sharp and Burrows 2007}.  The list
includes 55,300 lines, but again no width data are available and it was
estimated. 

\end{itemize}

\subsection{Line By Line Calculations}

As noted previously, we compute the opacity on a fine, fixed wavenumber grid.  We add the opacity arising from  each individual molecular line (hence, `line by line') using
a program that takes information for each molecular absorber from a database that contains line
strengths and positions, the lower energy level and line broadening
information. The line profiles are generated from the line database
with a Voigt profile algorithm; at higher pressures the profile
are essentially Lorentzian. We neglect the problem of how to treat the shape
of the far line wings, where it is known \cite{Levy et al. 1992} that the line shape
should eventually become sub-Lorentzian. In most cases actual measurements, particularly at the higher pressures,
 are lacking and in many cases only
a few theoretical predictions are available for selected bands.
We make no attempt to simulate the
specialized line shapes that are appropriate in the far IR and microwave
regions \cite[see][for a discussion in the context of Jupiter's deep atmosphere]{de Pater et al. 2005}. The line shapes in these regions are expected to be asymmetric,
but because of the large overlap of the low and high frequency wings
due to the high density of lines, the effects due to the deviation
of the line shape from a Lorentzian will tend to average out and should
not cause a significant change in the integrated values of the mean
opacities.

Likewise we neglect $\rm{\chi}$ factors \cite{Levy et al. 1992} which describe the deviation of the line wings
from a pure Lorenztian form. This includes effects such
as line mixing and line narrowing, for example. Some information
is available for selected species on the deviation of line shapes
from a Lorentzian, but in many cases these studies have covered only
a single band of a given molecular species, and the broadening agent
was usually some mixture of $\rm N_2$  and $\rm O_2$  instead of $\rm H_2$
and He which would be the appropriate choice for brown dwarfs and 
giant planets. After conducting a number of tests, we found
that the inclusion of a $\chi$ factor for $\rm H_2O$, a major source
of opacity, had no substantive effect on the Rosseland mean results.
Considering all the other sources of uncertainty, we did
not include $\rm{\chi}$ factors in this study.

The collisional (pressure broadened) line widths for several of the species
used in our opacity calculations are not well known. In general, there have
been no measurements reported in the literature and in many cases the main
information about the line positions and intensities 
that we use comes from theoretical predictions. One could estimate, in principle,
the line widths in comparison to other molecules by examining the relative values
of the molecular polarizabilities \cite{Hirschfelder et al. 1954, CRC Handbook 2000-2001}.
Unfortunately, even this information is not
available to us for many species of interest. It might be possible to again estimate
some of these polarizabilities by quantum mechanical calculations but again this data
is not currently available. For the one case that can be compared to measurements,
namely $\rm H_2S$, the large relative value of the static polarizability compared to other molecules
in our list does correlate well 
with the rather large line widths actually measured \cite{Kissel et al. 2002}.
Currently, we simply use estimates for the line widths of $\rm TiO$ and the metal
hydrides that are $\sim 25 - 50\%$ larger than the line widths for other molecules that are
earlier in the periodic table and generally  have smaller effective radii when they are
formed into molecules. 

As a check on the effect of uncertainties in the line broadening for the various
metal compounds, we calculated a set of $\rm TiO$ opacities with twice the
assumed value of the pressure broadening widths. A direct comparison shows
no significant effect on the overall mean opacity. The total value of the
absorption is conserved over the line profile as it should be, assuming that
the line wings are allowed to extend to larger values at higher pressures.
The overall effect of broader lines is to smooth out the central peaks of the absorption
lines and to fill in the low points in the far wings. Since the doubled 
widths approach the largest measured values for any molecule in our study ($\rm H_2S$),
it is apparent that line width uncertainties of a factor of 2 are not a significant
source of error in our calculations.

The remaining line broadening parameters were taken from the literature when
available. Available information can include actual measurements or
theoretical predictions.
In a few cases, no experimental or theoretical data of any kind for
$\rm H_2$ and He broadening was available so estimates of the line
width were made. In several cases data was available for broadening
by $\rm H_2$ and He and this was used. The assumed line broadening
parameters were scaled by pressure and an assumed temperature dependence
for each $(T, P)$ combination. This temperature dependence could come
from actual measurements or from theory. In practice, the laboratory measurements
do not cover a very large range in $T$, but this is usually all that
is available \cite{Homeier 2005}. In contrast, the Van der Waals theory of broadening
by foreign gases, commonly used in stellar atmospheres calculations,
predicts the same temperature dependence for all lines irrespective
of their angular momentum quantum numbers and the identity of the foreign broadener is usually
assumed to be hydrogen atoms.

The scaling of the line width linearly with pressure ignores the problem
of what happens to the line width at very high pressures, as a hard
sphere cutoff to the pressure scaling should exist. This is related
to the value of the second and higher virial coefficients in the
equation
of state for the gas. Since reliable experimental data on collision
cross sections as derived from viscosity and diffusion data are only
available for a few combinations of species and broadener, it is
difficult
to validate the theory under the physical conditions that apply to
the astronomical case for more than a few molecule-broadener
combinations.
In particular, in the astronomical case the highest pressures are
associated with the highest temperatures which is just the region
where the parameters that determine the equation of state are the
most uncertain.

  \subsection{Collision Induced Absorption due to $\rm{H_{2}-H_{2}}$,
$\rm{H_{2}-He}$ and $\rm{H_{2}-H}$}

Collision induced absorption produces a broad continuum that sculpts
the foundation of ultracool dwarf spectra.
Our source for the subroutines to calculate the collision induced
absorption comes from the recent work of A. Borysow and her
collaborators
\cite{Borysow 2002, Borysow et al. 2000, Borysow et al. 1997, Birnbaum et al. 1996, Zheng & Borysow 1995, Borysow 1992, Borysow & Frommhold 1990, Borysow et al. 1985}. We
have used the latest available
versions of all programs to compute the CIA absorption due to $\rm
H_2-H_2,$ $\rm H_2-He$ and $\rm H_2-H$
collisions. The FORTRAN programs and opacity tables available on
Borysow's web page\footnote{http://www.astro.ku.dk/\textasciitilde{}aborysow/} were  used
to construct tables that
represent the absorption by a mixture of {}``normal'' (3:1) $\rm H_2$.
This can be contrasted with an ``equilibrium'' mixture where the
ratio of the ortho and para forms of $\rm H_2$ is 1:1. At the high
temperatures of these objects a normal distribution would be expected.
This topic is discussed more thoroughly in Massie and Hunten (1982)
and Carlson et al. (1992). In any case, the difference in
the results between the two cases is only significant at the lowest
temperatures and at the low frequency end of the spectrum. This would
lead to small changes in the results of a few percent in these cases,
well within the other sources of uncertainty in this problem.

\subsection{Opacity from Alkali Atoms: Na, K, Cs, Rb and Li}

The importance of alkali atoms  to atmospheric opacity in cool
substellar objects was first
recognized from the influence of these atoms of the far red spectra of
T dwarfs \cite{Burrows et al. 2000}.  These pressure broadened lines,
particularly of Na and K, are major opacity sources over certain spectral ranges, temperatures, and
densities, particularly 
above the Na and K condensation temperatures (about 600 K, Lodders (1999a)).
Although the exact form of the line shapes of the alkali atoms due
to pressure broadening by $\rm{H_{2}}$ is still not completely
understood, it is important to include it. We use a computer
code  \cite{Burrows et al. 2000}  kindly provided to us by A. Burrows 
to generate line profiles for atomic lines of neutral alkali atoms
using a line width parameter setting of 1.0 (as defined by those
authors).

Burrows and Volobuyev (2003) and Allard et al. (2003) have further modeled
the pressure broadened alkali lines.  Future improvements in understanding
of the alkali line widths will certainly impact spectral modeling of ultracool dwarfs
and may impact the mean opacities.  The importance of the alkali elements to
 the mean opacities is discussed in Section 4.

\subsection {Other Opacity Sources}

Several other opacity sources are included in our calculation.
Bound-free absorption by H and $\rm H^{-}$ and free-free absorption by
H, $\rm H_2$,  $\rm H_{2}^{-}$ and $\rm H^{-}$ (see Lenzuni et al.
1991) were added using algorithms provided by Tristan
Guillot \cite{Guillot 1999}.
Rayleigh scattering from $\rm H_{2}$ and Thompson scattering are also
included in the Rosseland mean following Lenzuni et al.
Opacity from electrons and H atoms does not provide  a large
contribution to the overall opacity below $\sim 2500\,\rm K$ but
becomes important at higher temperatures relevant to M dwarf
atmospheres.

The opacity from the more abundant atomic species are not important in our case because our high temperature cases also involve high pressure.  Since we are concerned with high gravity objects the fractional abundances of atoms like Fe, Mg, Si, and Al never exceed $8\times10^{-5}$ and are usually much
lower.  Furthermore the most important lines for these atoms are in the UV where there is little flux over our temperature regime.

\section{Chemical Equilibrium Calculations}

The thermochemical equilibrium abundances used in the opacity models
were computed with the CONDOR code described and applied to brown dwarf
studies in several papers (e.g., Fegley \& Lodders 1994, 1996, Lodders
1999a, 2002, Lodders \& Fegley 2002). Brief overviews about the gas
and cloud chemistry in substellar atmosphere can be found in Lodders
(2004) and Lodders \& Fegley (2006).  Here we summarize this work and
highlight important issues for the problem at hand.

\subsection{Gas and Condensate Chemistry Computations}

The CONDOR code simultaneously computes the chemical equilibrium compositions for
more than 2200 gases (including ions) and more than 1700 solids and liquids of all naturally
occurring elements by considering dual constraints of mass balance and chemical equilibrium.
The major thermodynamic data sources are given in Fegley \& Lodders (1994), and data are
frequently updated (see e.g., Lodders 1999b, 2004). Note that some frequently used compilations of thermodynamic properties and/or polynomial fits contain errors that also affect compounds important for brown dwarf chemistry (see Appendix 1 in Lodders 2002). Therefore chemical equilibrium species and abundances computed by other groups could be different than our results if erroneous thermodynamic data were used by them, which can also introduce differences in calculated opacities.

The CONDOR code uses elemental abundances, total pressure and
temperature as inputs. We use the solar system abundances in Lodders
(2003), uniformly enhanced or depleted to model metallicity effects
where appropriate. Equilibrium compositions considering cloud
condensate formation for solar elemental abundances were computed for
324 pressure-temperature sample points in a grid ranging from 50 to
4000 K and $\log P({\rm dyne\, cm^{-2}}) = 2$ to +8.5 which spans characteristic
conditions in the atmospheres of low mass objects. Similar calculations
were done for other metallicities; here we only include results for
metallicities of 2 and 1/2 times solar ($\rm [M/H] = \pm 0.3$).

Despite the plethora of gas species present in the thermochemical
calculations, only a few compounds are abundant or major opacity
sources, and the subset of the gases selected here for constructing
the opacity tables are discussed in Section 2.
Compared to previous solar elemental abundance compilations, the more
recent data include significant downward revisions in the C, N, and O
abundances around 20\% to 40\%,
which have consequences for the abundances of important opacity sources
such as methane, CO, water, and ammonia in substellar atmospheres. The
new lower C, N, and O abundances resemble a decrease in CNO
metallicity, and a detailed discussion of how metallicity affects the
$\rm CH_4/CO$ and $\rm NH_3/N_2$ equilibria is given in Lodders \& Fegley
(2002). For example, one important consequence of the lower C, N, and O
abundances is that the methane to CO as well as the $\rm NH_3$ to $\rm
N_2$ conversions are shifted towards higher temperatures (at constant
total pressure or gravity), and are shifted towards lower total
pressure (at constant temperature).

The thermochemical calculations also include results for deuterium.
However, in the opacity modeling for brown dwarfs, the deuterium
abundance was set to zero because it is assumed that all D is destroyed
in objects more massive than 13 Jupiter masses. Thus, the opacities due from
HDO and $\rm CH_{3}D$ are not included. However, in lower mass objects,
the HDO and $\rm CH_{3}D$ opacities must be included and these were
calculated using $\rm D/H = 1.94 \times 10^{-5}$ \cite{Lodders 2003}. However, the additional
opacity from HDO and $\rm CH_{3}D$ generally has only a very marginal
effect on the overall results.

\subsection{Condensate treatment}
The CONDOR code has two principle pathways to treat condensate
formation, depending on the desired application. Under ideal
equilibrium conditions in protoplanetary disks or stellar winds,
gas-solid equilibria are maintained within a cooling gas and therefore
high temperature condensates that formed first from a cooling gas
(primary condensates)
still can react with the gas to form secondary condensates at lower
temperatures. A well-known gas-solid reaction is the reaction of
primary iron metal with $\rm H_2S$ gas to secondary troilite (FeS) at
low temperatures (see below). However, such reactions may not apply to
substellar
atmospheres where cloud layer formation prevents the formation of
secondary condensates (see Lodders \& Fegley 2002). If a primary
condensate settles from the gas into a cloud layer (sometimes called
rainout),
the primary condensate becomes depleted in the atmosphere above its
cloud and cannot participate in reactions that are thermochemically
favorable at the cooler temperatures above the cloud. For example, if
iron metal condenses and settles into a cloud, the reaction of iron
metal with $\rm H_2S$ to secondary troilite cannot happen. This has
important consequences for the gas chemistry because troilite formation
would remove all $\rm H_2S$ (essentially all the atmospheric sulfur
inventory) at the low temperatures where troilite would be stable.
However, if the secondary troilite does not form, $\rm H_2S$ remains in
the atmosphere until $\rm NH_4SH$ condenses, which only happens at $T
<<300\,\rm K$. The {\it Galileo} entry probe mass spectrometer detection of
$\rm H_2S$ at about three times the solar $\rm S/H_2$ ratio in
Jupiter's atmosphere below the $\rm NH_4SH$ cloud level (Niemann et al. 1998)
shows that the cloud layer condensation approach works well in giant
planet atmospheres such as in Jupiter (see also Lodders \& Fegley 2002,
Visscher et al. 2006).

The computations with cloud settling thus require the knowledge of
which condensates are primary (formed directly from the gas) and which
ones are secondary (formed by gas-solid reactions). This is easily
gauged from the equilibrium calculations by checking the distribution
of an element between condensates and gas. For example, under typical
total pressure conditions (e.g., $>10^5\,\rm dyne\,cm^{-2}$) in substellar atmospheres,
iron starts to condense as metal at temperatures above 1600 K, and the
metal settles into a cloud layer with a base at the temperature level
where metal condensation starts. At the cloud base, iron metal is in
evaporation-condensation equilibrium with the hotter atmosphere below.
Above the cloud base, the iron gas abundance is determined by the iron
vapor pressure of the metal cloud. The iron vapor pressure drops
exponentially with the decreasing temperatures above the cloud, and
mass balance dictates that more iron is in the condensates, and less
iron gas remains in the atmospheric gas. Typically it takes
about a $100 - 200\,\rm K$ drop from the condensation temperature
(i.e., the temperature where the condensate appears first from a
cooling gas) to the temperature where $>99\%$ of an element is removed by
its condensate. This is the normal vapor-pressure driven condensation
process. The only special consideration here is that the condensate
settles so that the amount of condensate becomes depleted at
atmospheric levels above the condensate cloud that forms from the
settled condensate particles. However, this does not affect the fact
that the gas abundances remain established by the vapor pressure over
the condensing solid or liquid because the vapor pressure over a
substance is independent of the absolute amount of the substance
present.

Troilite formation would only occur at temperatures below about 700 K,
which is significantly lower than the temperatures at which iron vapor
pressures are so low that essentially all iron is condensed. The
gas-sold reaction for troilite formation requires the presence of iron
metal, but at the temperature level where FeS becomes stable, the iron
metal condensate has settled out at greater depth in low mass object
atmospheres, and thus, the secondary troilite cannot form.

Note that the thermochemical equilibrium calculations of the gas
abundances do not require information about the cloud particle sizes,
their settling efficiencies, or the vertical extent of the cloud layer;
nor can these calculations provide this information without further
model assumptions. These parameters are not needed if we are only
interested in gas opacities, however, these parameters are needed if
condensate opacities are to be considered, which requires inclusion of
a detailed cloud model (see e.g., Ackerman and Marley 2001, Marley et
al. 2002).

Other observations also demonstrate that other refractory elements are
depleted by condensate cloud formation at high temperatures deep in the
atmospheres of low-mass objects. For example, the gradual disappearance
of TiO and VO bands in the optical spectra of L dwarfs requires
condensation of Ca-titanates into which VO dissolves (Fegley \& Lodders
1994, 1996, Lodders 1999a, 2002). The absence of silane ($\rm SiH_4$)
indicates deeply seated silicate clouds and the presence of volatile
germane ($\rm GeH_4$) in the atmospheres of Jupiter and Saturn shows
that germanium did not condense into iron metal at low temperatures (as
it did in the solar nebula, see Lodders 2003) because the metal settled
(Fegley \& Lodders 1994). The presence of monatomic K gas in the
atmospheres of T dwarfs requires that the refractory
feldspar anorthite has settled into a deep cloud because otherwise, the
more volatile K could dissolve in it at lower temperatures, (e.g.,
 Lodders 1999a, Burrows et al. 2000, Geballe et al. 2001, Marley et al. 2002).

In summary, the major difference for {\it gas} chemistry between the ``nebula" and ``cloudy" cases is that the gas abundances of elements forming condensates by gas-solid reactions will be lower in the ``nebula" models than in the ``cloudy" models at the temperatures and total pressures where such secondary condensation is expected. In the cloudy models, the secondary condensates cannot form because the solid required for the gas-solid reaction is not present; e.g.,  anorthite is required for Na and K condensation in solid-solution. Therefore, the {\it gas opacities} will be different for the nebula and cloudy models.  As a practical matter, the biggest impact on the opacities arises from the timing of removal of gaseous alkali species.  In the ``cloudy'' chemisty, Na and K persist to lower temperatures than would otherwise be the case.  The spectra of T dwarfs confirms that in fact the ``cloudy'' chemistry is the correct choice for atmospheres.  Since we neglect grain opacity, the differences in the timing and arrival of the condensates themselves does not impact our opacity calculation.

\section{Results}

\subsection{Mean Opacity Tables}

Given the opacities described in Section 2 and the elemental and
molecular abundances described in Section 3, we computed Rosseland and
Planck mean opacities for three metallicities ($\rm [M/H]= -0.3, 0.0,
+0.3$).
The integration
over frequency was carried out using a grid of equally spaced wavenumber
points. The frequency spacing was based on the temperature and pressure
of each layer, such that the spectral resolution was always fine (1/4 of a line width
or $0.5\,\rm cm^{-1}$, whichever is less) compared to a
typical
line profile under those conditions. Spacings are as small as  $4\times10^{-3}\,\rm cm^{-1}$ in some
instances.  Such a fine grid is usually not
required for mean opacity calculations, but we also use the same grid
for high resolution spectral modeling.

Our grid of 324
$(P,T)$ pairs ranges from 75 to $4000\,\rm K$ and  $3\times 10^{2}$ to $3\times 10^8\,\rm dyne\,cm^{-2}$ (or $3\times 10^{-4}$ to 300 bar).  We do not compute opacities at high pressure and low
temperature or high temperature and low pressure, as such points are not reached by
brown dwarf and giant planet atmosphere models.  The table spacing is not uniform in order
to better sample important processes, including water condensation.
 Figure 2 illustrates the pressure-temperature domain over
which we computed opacities as well as the magnitude of the Rosseland
mean opacities for a solar composition gas.  This figure also shows
temperature-pressure profiles  computed for Jupiter and a
variety of other substellar objects.  
Tables I, II, and III provide Rosseland and Planck mean opacities as a function
of pressure and temperature for the three metallicities we consider on this grid.  
The complete tables may be found in the on line
supplement to this paper.  In the tables opacities have been converted to units
of $\rm cm^2/g $  by calculating
the mean molecular weight (for gaseous species only) at each temperature-pressure level, for each
chemical composition.

Because of our choice of a uniform pressure-temperature $(P, T)$ grid, our data are not
on a uniform mass density ($\rho$) grid.
For easier comparison with earlier work, we also interpolated to a set of
constant densities for graphical purposes.  Figures 3  illustrates the
Rosseland mean opacities along profiles of constant density.  We note that
such an interpolation on occasions crosses chemical equilibria and condensation
boundaries, which produces some kinks in the interpolated data shown on the figure.  

\subsection{Opacity from Alkali Atoms: Na, K, Cs, Rb, and Li}

Perhaps the greatest difference from previous tabulations of mean
opacities arises
from the inclusion of the pressure-broadened lines of the alkali
elements, particularly
sodium and potassium.  These molecules, with their large absorption
cross sections
in the far red and very near infrared (Burrows et al. 2000) fill what
would otherwise
be a region of fairly low opacity.   Figure 4 compares the summed
opacity as a function of wavenumber at 1400 K and $10^6\,\rm dyne\,cm^{-2}$ in our baseline case with
a calculation that neglects the alkali opacity.  The substantial role of
the alkali opacity to the total summed opacity is unmistakable above about
$10,000\,\rm cm^{-1}$.  The influence
of the alkali opacity on the total gaseous mean opacity is illustrated in Figure 5
which compares the Rosseland
mean opacity with and without the contribution of alkali metals
at several densities.  It is clear that at the higher
densities the  alkali opacity substantially fills in the opacity
minimum from about 1000 to 1500 K.  Differences at lower densities are
slight since the pressure-broadened lines play a much smaller role.

Guillot et al. (1994a, b) predicted that low opacity around 1,000 to 2,000 K in 
Jupiter's deep atmosphere would lead to the formation of a radiative region 
within what was then expected to be the fully convective deep interior of Jupiter.
In Figure 2 it can be seen by extrapolation that Jupiter's deep adiabat indeed passes through
a trough in opacity in this temperature range.  We now understand that the opacity in this local  
minimum region--owing to the contribution of pressure-broadened alkali opacity--is much 
larger than was considered by Guillot et al. (1994a,b) using the opacities available at that time.
In a reevaluation of their earlier work, Guillot et al. (2004) indeed found that the alkali opacity is sufficient to prevent substantial
energy transport by radiation in Jupiter's interior.  The Rosseland opacities we report here are consistent with
those used in the latest work by Guillot and collaborators.  Thus, despite the intrinsic uncertainty
in the alkali line widths at high pressure, a thin radiative shell within the interior
of Jupiter  need not be considered in the construction of interior models.  This  removes one source
of uncertainty in the construction of evolution and interior models of Jupiter.

As noted in \S2.4, the pressure broadened line shape for the alkali metals
remains uncertain, particularly
at high pressure.  In the future, it may be possible to obtain
experimental data on the line absorption coefficients of pressure
broadened alkali lines so that the effects of their shapes on the total
opacity can be quantified to replace the semi empirical profiles \cite{Johnas et al. 2006}.

\section{Discussion}

The opacities and chemical equilibrium calculations described here are examples of
the current state of the art for understanding giant planet and brown dwarf atmospheres.
Nevertheless, a number of uncertainties remain, particularly in the treatment of
the opacities at high temperatures.   Because it is a dominant source
of opacity, of course water is of special concern,
but  comparisons between the most
recent theoretical line lists  suggest that, at least the the temperatures considered here,
the line list is in reasonably good shape. 
Further work on calculating an updated
version of the the water spectrum
is currently underway. Missing opacity at high temperatures is certainly 
a problem in even greater
measure for most of the other molecular opacities, particularly methane, but these
other molecules are generally less important than water to the mean opacity. In fact, only water, CO, TiO, and to a lesser
extent VO have a significant contribution from hot bands while $\rm
CH_4$ and, to a lesser extent, $\rm H_2S$ contain lines originating from
higher rotational quantum values which have been predicted from a mixture of theory
and available experimental data. 

For comparisons with spectra of ultracool dwarfs, however,  the lack of adequate
high temperature opacities for methane and ammonia is an important
limitation (e.g., Leggett et al. 2007). 
Improving this situation will require in the case of methane a
substantial theoretical and computational effort. Newer data are under
development for water by David Schwenke and for CO by one of us (R.F.) that
would allow a recomputation of the line lists. In the case of water
this could lead to a better representation of the opacity at the highest
temperatures, while any changes in the CO database will probably be
less substantial. 

Because the calculations  presented here do not include condensates as an opacity source
a direct comparison with earlier work that includes grains is difficult.
In our brown dwarf and extrasolar planet modeling calculations 
cloud opacity is computed from the local description of the atmospheric
structure, rather than relying on a pre-computed table.  
 Because these additional
sources of opacity may appear at different pressures and temperatures
in a series of models depending on the assumptions built into the
calculation, it is not possible to give a general set of results that
include solid material. Any such tables must be regarded with some
caution as condensate size and abundance depends on other parameters,
including the convective velocity, and no single prescription as a function of only
density and temperature can be given.  In fact the grain opacity plays
an important role in the gaseous accretion of giant planets by the core accretion
mechanism \cite{Hubickyj et al. 2005}.

We show in Figure 6 a
comparison with recent grain free calculations  \cite{Ferguson et al. 2005}.
This work uses updated atomic abundances to compute mean opacities at 
relatively low densities relevant to circumstellar disks.  As such the region
of overlap in density and temperature is relatively small.  In the overlapping region,
however, the correspondence is reasonably good.  Since this is a low density
region, the effect of the alkali opacity is negligible and is not a factor.   

\section{Conclusion}

The tables presented here and in the on line supplement to this paper provide Rosseland and Planck mean opacities for three elemental compositions relevant to the study of ultracool dwarfs and extrasolar giant planets.  We have also described the databases for line transitions and our approach to computing the line broadening as well as the chemical equilibrium calculation.  Future improvements
in the molecular opacities--particularly at high temperature--will certainly improve the quality of model spectra for the comparison with astronomical data.  Barring the addition of substantial new
opacity sources or further updates to the solar abundance of the elements, we do not expect to see
significant changes to the Rosseland and Planck mean opacities reported here.

\acknowledgements
We thank Tristan Guillot for providing his hydrogen opacity routine and for
helpful advice, Jonathan Fortney for help with
figures and formatting, Didier Saumon for helpful conversations, 
Adam Burrows for use of his alkali atom opacity code, and the referee for suggestions which improved the manuscript.  
We also received generous support from David Schwenke and his collaborators.
He has provided extensive calculations and information concerning the quantum
mechanical modeling of various spectra.
R.F. acknowledges support from NASA grant NAG5-4970, M.M. acknowledges
support from the NASA Office of Space Sciences. Work by K.L. is supported by NSF grant AST0406963 and NASA grant NNG06GC26G.

\expandafter\ifx\csname natexlab\endcsname\relax\def\natexlab#1{#1}\fi

\clearpage
\begin{deluxetable}{ccccc}
\center
\tablecolumns{5}
\tablewidth{0pc}
\tablecaption{Mean Opacities for $\rm [M/H]=0.0$}
\tablehead{
\colhead{T (K)} &\colhead{$\rm P(dyne\,cm^{-2}$)} & \colhead{$\rho(\rm g\, cm^{-3})$} & \colhead{$\kappa_{\rm R}(\rm cm^2\,g^{-1})$} & \colhead{$\kappa_{\rm P}(\rm cm^2\,g^{-1})$} } 
\startdata
 75 & 3E+02 & 1.1277E-07 & 2.5619E-06 & 7.1083E-06 \\
75 & 3E+03 & 1.1277E-06 & 2.5589E-05 & 6.4309E-05 \\
75 & 1E+04 & 3.7591E-06 & 8.5261E-05 & 2.1238E-04 \\
75 & 3E+04 & 1.1277E-05 & 2.5571E-04 & 6.3555E-04 \\
75 & 1E+05 & 3.7591E-05 & 8.5211E-04 & 2.1167E-03 \\
75 & 3E+05 & 1.1277E-04 & 2.5557E-03 & 6.3485E-03 \\
75 & 1E+06 & 3.7591E-04 & 8.5180E-03 & 2.1160E-02 \\
75 & 3E+06 & 1.1277E-03 & 2.5553E-02 & 6.3478E-02 \\
75 & 1E+07 & 3.7591E-03 & 8.5176E-02 & 2.1159E-01 \\
100 & 3E+02 & 8.4584E-08 & 4.5393E-06 & 2.4757E-02 \\
100 & 3E+03 & 8.4583E-07 & 3.9962E-05 & 2.5407E-03 \\
100 & 1E+04 & 2.8193E-06 & 1.2854E-04 & 1.0837E-03 \\
100 & 3E+04 & 8.4582E-06 & 3.7709E-04 & 1.0589E-03 \\
100 & 1E+05 & 2.8193E-05 & 1.2345E-03 & 2.5780E-03 \\
100 & 3E+05 & 8.4582E-05 & 3.6583E-03 & 7.3903E-03 \\
100 & 1E+06 & 2.8193E-04 & 1.2104E-02 & 2.4401E-02 \\
100 & 3E+06 & 8.4582E-04 & 3.6260E-02 & 7.3044E-02 \\
100 & 1E+07 & 2.8193E-03 & 1.2088E-01 & 2.4334E-01 \\
100 & 3E+07 & 8.4582E-03 & 3.6261E-01 & 7.2982E-01 \\
\enddata
\tablecomments{Complete table on line.}
\end{deluxetable}
\clearpage
\begin{deluxetable}{ccccc}
\center
\tablecolumns{5}
\tablewidth{0pc}
\tablecaption{Mean Opacities for $\rm [M/H]=+0.3$}
\tablehead{
\colhead{T (K)} &\colhead{$\rm P(dyne\,cm^{-2}$)} & \colhead{$\rho(\rm g\, cm^{-3})$} & \colhead{$\kappa_{\rm R}(\rm cm^2\,g^{-1})$} & \colhead{$\kappa_{\rm P}(\rm cm^2\,g^{-1})$} }
\startdata
75 & 3E+02 & 1.1313E-07 & 2.5527E-06 & 7.6946E-06\\
75 & 3E+03 & 1.1313E-06 & 2.5496E-05 & 6.4814E-05\\
75 & 1E+04 & 3.7710E-06 & 8.4951E-05 & 2.1216E-04\\
75 & 3E+04 & 1.1313E-05 & 2.5478E-04 & 6.3330E-04\\
75 & 1E+05 & 3.7710E-05 & 8.4894E-04 & 2.1074E-03\\
75 & 3E+05 & 1.1313E-04 & 2.5460E-03 & 6.3191E-03\\
75 & 1E+06 & 3.7710E-04 & 8.4845E-03 & 2.1060E-02\\
75 & 3E+06 & 1.1313E-03 & 2.5452E-02 & 6.3178E-02\\
75 & 1E+07 & 3.7710E-03 & 8.4837E-02 & 2.1059E-01\\
100 & 3E+02 & 8.4848E-08 & 4.5234E-06 & 2.4853E-02\\
100 & 3E+03 & 8.4846E-07 & 3.9824E-05 & 2.6108E-03\\
100 & 1E+04 & 2.8282E-06 & 1.2808E-04 & 1.1500E-03\\
100 & 3E+04 & 8.4846E-06 & 3.7575E-04 & 1.1219E-03\\
100 & 1E+05 & 2.8282E-05 & 1.2300E-03 & 2.6327E-03\\
100 & 3E+05 & 8.4846E-05 & 3.6443E-03 & 7.4225E-03\\
100 & 1E+06 & 2.8282E-04 & 1.2056E-02 & 2.4353E-02\\
100 & 3E+06 & 8.4846E-04 & 3.6115E-02 & 7.2772E-02\\
100 & 1E+07 & 2.8282E-03 & 1.2042E-01 & 2.4226E-01\\
100 & 3E+07 & 8.4846E-03 & 3.6123E-01 & 7.2650E-01\\
\enddata
\tablecomments{Complete table on line.}
\end{deluxetable}

\begin{deluxetable}{ccccc}
\center
\tablecolumns{5}
\tablewidth{0pc}
\tablecaption{Mean Opacities for $\rm [M/H]=-0.3$}
\tablehead{
\colhead{T (K)} &\colhead{$\rm P(dyne\,cm^{-2}$)} & \colhead{$\rho(\rm g\, cm^{-3})$} & \colhead{$\kappa_{\rm R}(\rm cm^2\,g^{-1})$} & \colhead{$\kappa_{\rm P}(\rm cm^2\,g^{-1})$} }
\startdata
75 & 3e+02 & 1.1260E-07 & 2.5662E-06 & 6.8130E-06\\
75 & 3e+03 & 1.1260E-06 & 2.5632E-05 & 6.4041E-05\\
75 & 1e+04 & 3.7535E-06 & 8.5404E-05 & 2.1245E-04\\
75 & 3e+04 & 1.1260E-05 & 2.5615E-04 & 6.3653E-04\\
75 & 1e+05 & 3.7535E-05 & 8.5363E-04 & 2.1208E-03\\
75 & 3e+05 & 1.1260E-04 & 2.5605E-03 & 6.3618E-03\\
75 & 1e+06 & 3.7535E-04 & 8.5345E-03 & 2.1205E-02\\
75 & 3e+06 & 1.1260E-03 & 2.5603E-02 & 6.3614E-02\\
75 & 1e+07 & 3.7535E-03 & 8.5343E-02 & 2.1205E-01\\
100 & 3e+02 & 8.4455E-08 & 4.5476E-06 & 2.4781E-02\\
100 & 3e+03 & 8.4453E-07 & 4.0027E-05 & 2.5126E-03\\
100 & 1e+04 & 2.8151E-06 & 1.2874E-04 & 1.0526E-03\\
100 & 3e+04 & 8.4453E-06 & 3.7770E-04 & 1.0280E-03\\
100 & 1e+05 & 2.8151E-05 & 1.2366E-03 & 2.5503E-03\\
100 & 3e+05 & 8.4453E-05 & 3.6650E-03 & 7.3736E-03\\
100 & 1e+06 & 2.8151E-04 & 1.2127E-02 & 2.4421E-02\\
100 & 3e+06 & 8.4453E-04 & 3.6329E-02 & 7.3175E-02\\
100 & 1e+07 & 2.8151E-03 & 1.2109E-01 & 2.4383E-01\\
100 & 3e+07 & 8.4453E-03 & 3.6325E-01 & 7.3143E-01\\
\enddata
\tablecomments{Complete table on line.}
\end{deluxetable}

\clearpage

\begin{figure}
\includegraphics[angle=270,scale=0.6]{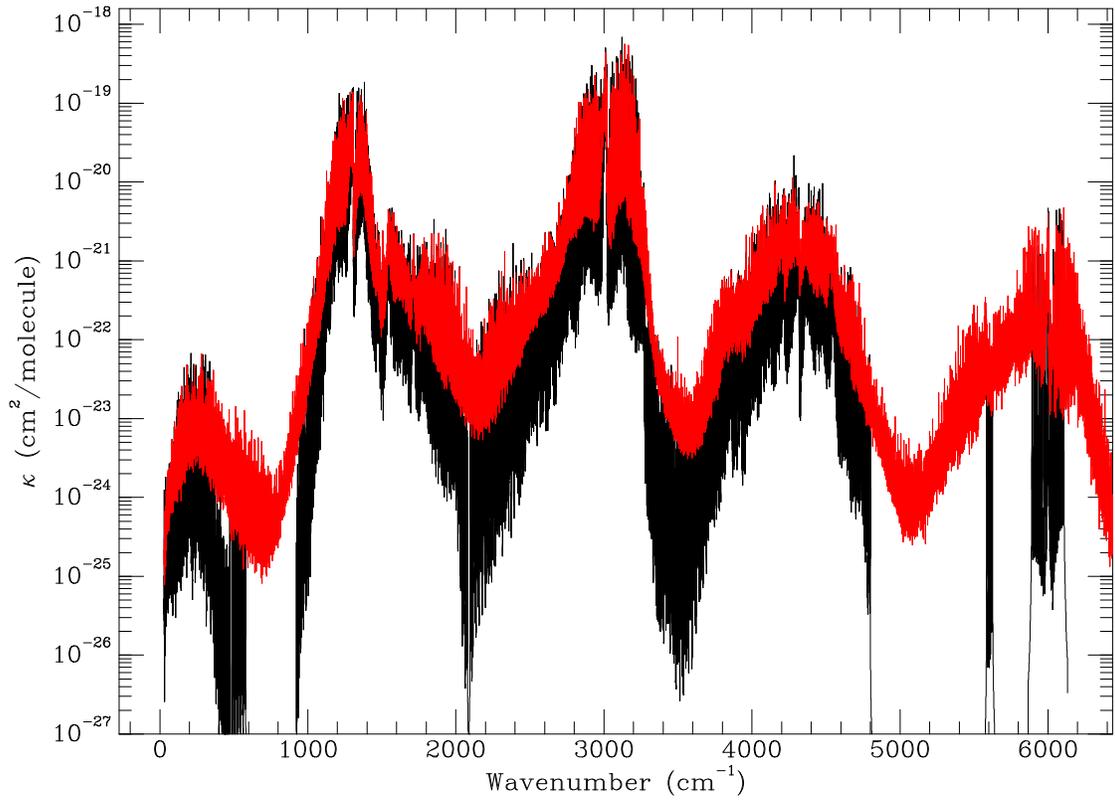}
\caption{Opacity $\kappa$ at $10^6\,\rm dyne\,cm^{-2}$ (1 bar) and 1000 K as a function of wavenumber for two opacity databases
of the $\rm CH_4$ molecule.  Black data are computed from the standard HITRAN database.  Red
curve is derived from our computational database, as described in the text.   }
\end{figure}

\begin{figure}
\includegraphics[angle=90,scale=0.7]{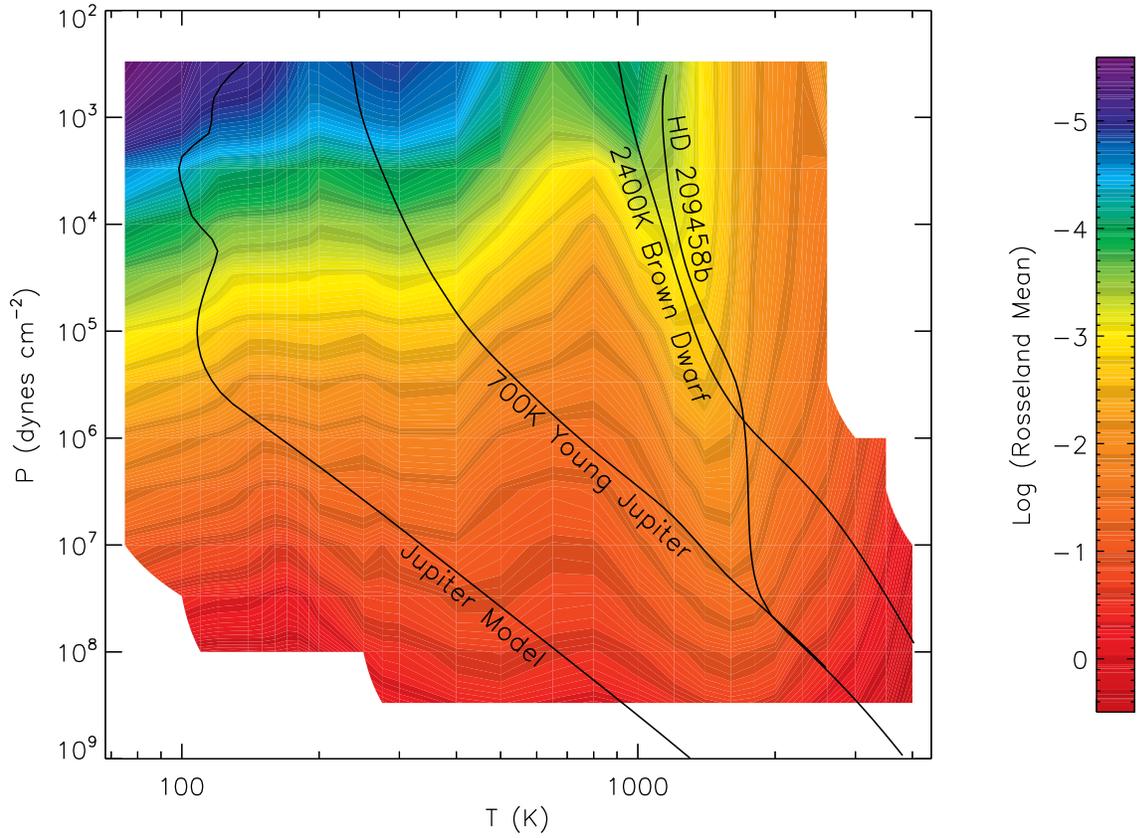}
\caption{Contour plot showing Rosseland mean opacity over the temperature-pressure domain
included in this study.  Model temperature-pressure profiles from several planets and a brown dwarf
are shown for comparison.   Temperatures along profile refer to the effective temperature; 
brown dwarf profile is for an object with surface gravity $\log g \rm (cm^2/sec) = 5$.}
\end{figure}

\begin{figure}
\includegraphics[angle=-90,scale=0.5]{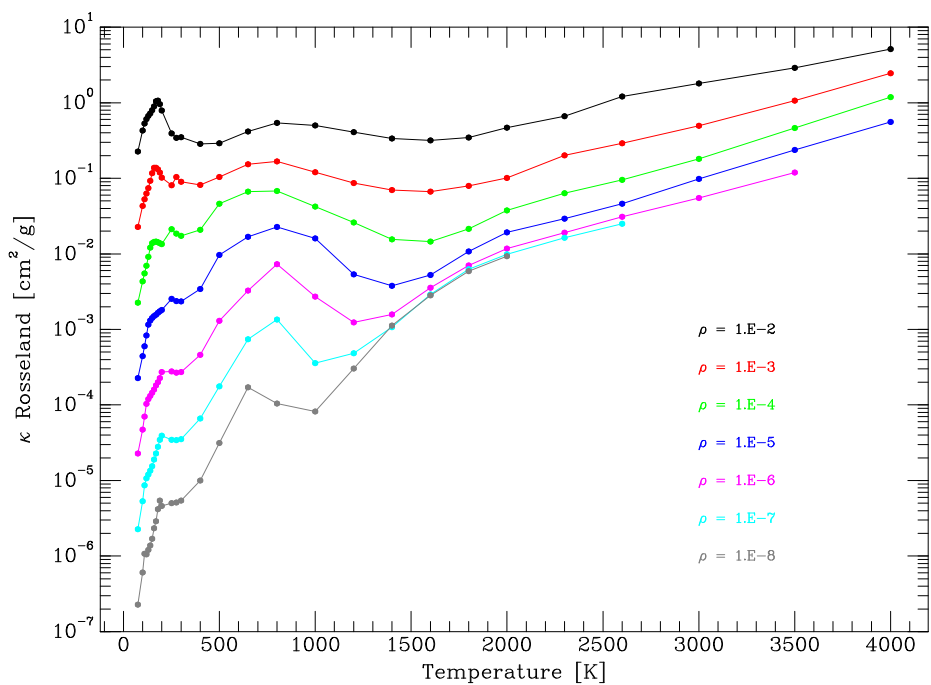}
\caption{Rosseland mean opacity at seven densities as a function of temperature.  Opacity data
at each $(\rho, T)$ pair is computed by interpolation within our standard grid.  Some discontinuities in this
figure arise from interpolation over chemical equilibria and condensation boundaries.  Densities range from $1\times10^{-8}$ (bottom) to 
$1\times10^{-2}\rm g/cm^3$.  }
\end{figure}

\begin{figure}
\includegraphics[angle=270,scale=0.6]{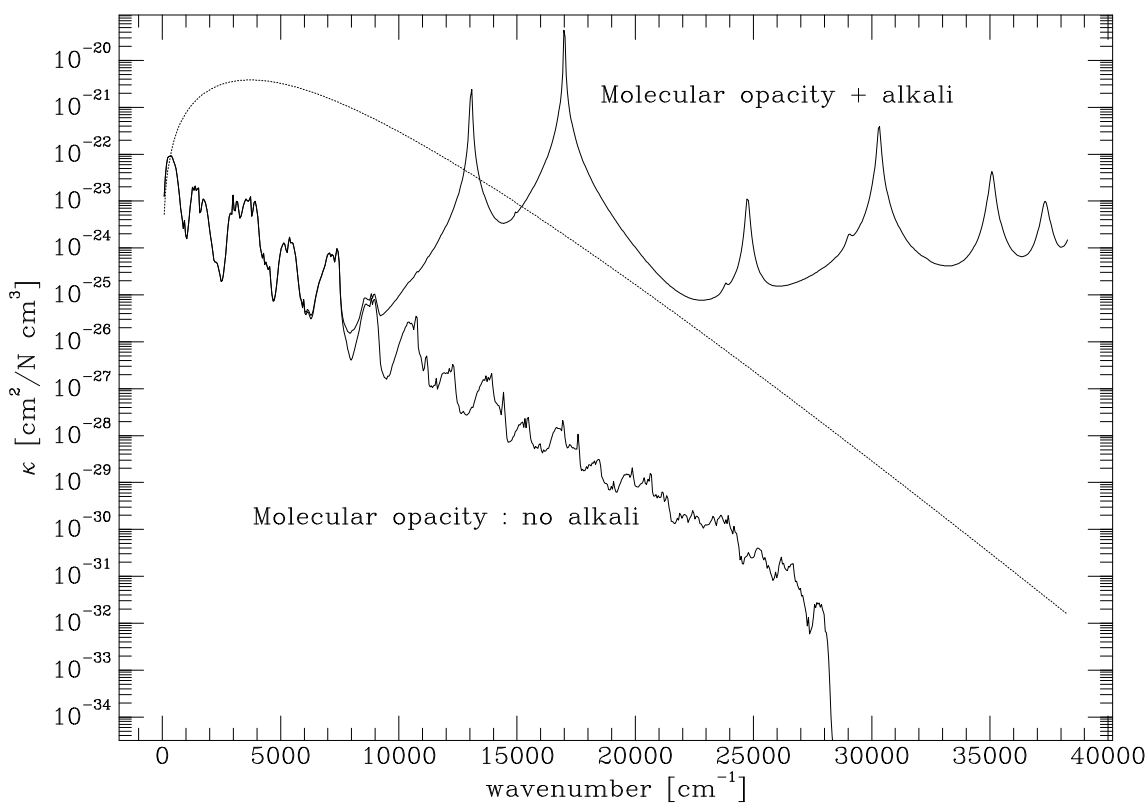}
\caption{Total opacity as a function of wavenumber at 1400 K and 1 bar.  Thin curve gives opacity without the
contribution of alkali atoms.  Thicker curve shows summed opacity including alkalis, computed using the 
theory of Burrows, Marley, and Sharp (1999).  Under these conditions the alkali opacity dominates at wavenumbers above
about $10,000\,\rm cm^{-1}$.   The derivative of the Planck function, $dB/dT$, which weights the
opacity in the computation of the Rosseland mean, is shown as as a dotted line on an arbitrary log scale.  }
\end{figure}

\begin{figure}
\includegraphics[angle=270,scale=0.6]{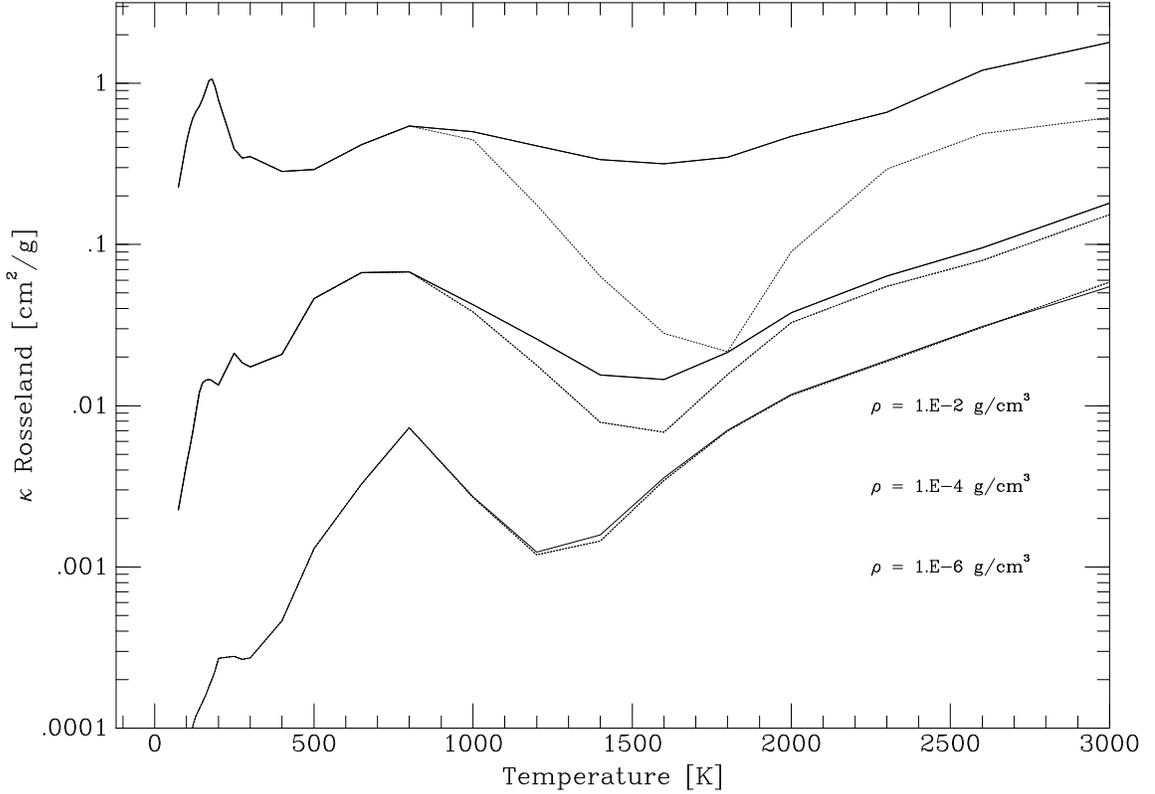}
\caption{Rosseland mean opacity at three densities as a function of temperature.  Opacity data
at each $(\rho, T)$ pair is computed by interpolation within our standard grid.  Some discontinuities in this
figure arise from interpolation noise.  Densities range from $1\times10^{-6}$ (bottom) to 
$1\times10^{-2}\rm g/cm^3$.  Solid line denotes our standard calculation, the dotted line is for a case
with the alkali opacity set equal to zero.  The alkali opacity substantially increases the mean opacity at
high densities and temperatures above about 1000 K.  Differences in the lowest density case plotted
are negligible. }
\end{figure}

\begin{figure}
\includegraphics[angle=270,scale=0.6]{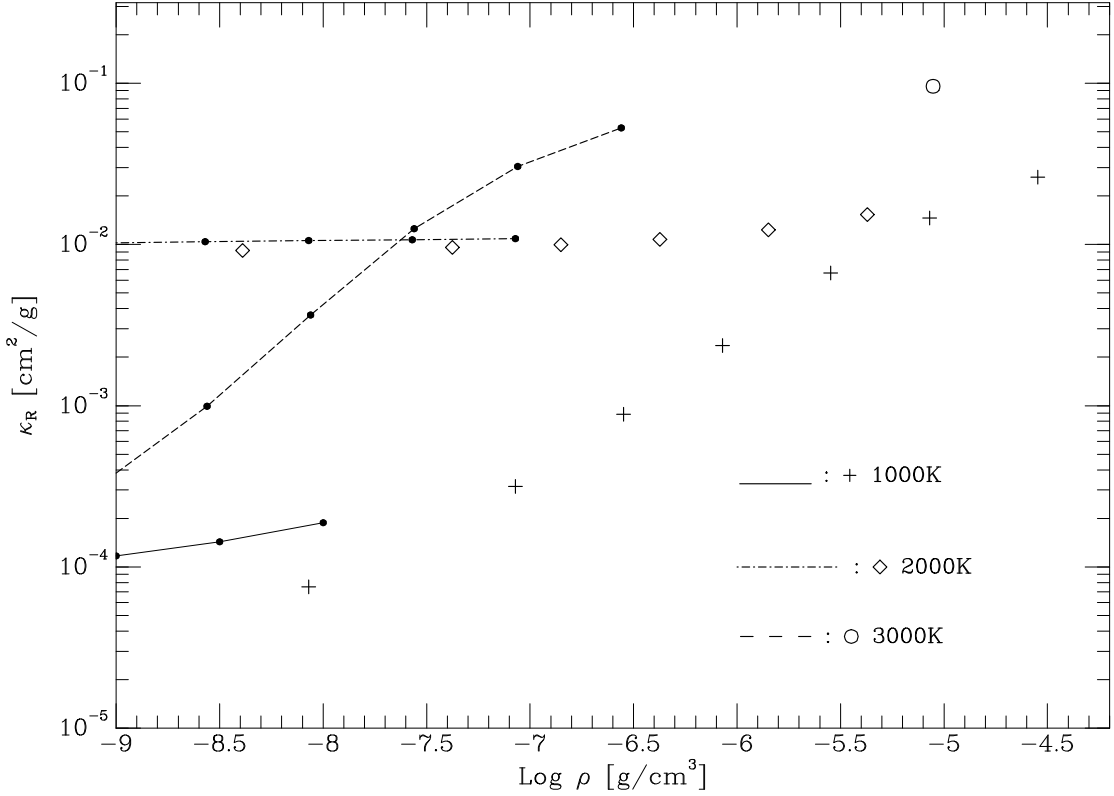}
\caption{Rosseland mean opacity as a function of density for three temperatures.  Solid and broken
lines are  the ``no grain'' opacity from
Ferguson et al. (2005) for three temperatures (specifically their case with $X=0.7$ and $Z=0.02$, filename=``cunha06.nog.7.02.tron'').  This opacity database is optimized for protostellar disks and thus there is relatively little overlap in density space with the cool atmospheric conditions we consider here.  The closest densities from our calculation  (including alkali opacity) are shown as isolated symbols.  There is excellent agreement at 2000 K.  The trends for 1000 and 3000 K suggest reasonably good agreement between the two calculations.}
\end{figure}

\clearpage

\end{document}